\documentclass[useAMS]{gGAF2e}

\newcommand{\lta}{\;
  \raise0.3ex\hbox{$<$\kern-0.75em\raise-1.1ex\hbox{$\sim$
  }}\;\hskip-2pt }
\newcommand{\gta}{\;
  \raise0.3ex\hbox{$>$\kern-0.75em\raise-1.1ex\hbox{$\sim$
  }}\;\hskip-2pt }
\newcommand{\apropto}{\;
  \raise0.3ex\hbox{$\propto$\kern-0.75em\raise-1.1ex\hbox{$\sim$
  }}\;\hskip-2pt }

\begin{document}
\doi{10.1080/03091920xxxxxxxxx}
 \issn{1029-0419} \issnp{0309-1929} \jvol{00} \jnum{00} \jyear{2009} 

\markboth{Dmitry Sokoloff and David Moss}{What can we say about seed fields for galactic dynamos?}

\title{{\textit{What can we say about seed fields for galactic dynamos?}}}

\author{DMITRY SOKOLOFF${\dag}$$^{\ast}$\thanks{$^\ast$Corresponding author. Email: d$_{-}$sokoloff@hotmail.com} and DAVID MOSS${\ddag}$
\vspace{6pt}\\\vspace{6pt}  ${\dag}$Department of Physics, Moscow
State University, Moscow, 119992, Russia\\ ${\ddag}$ School of
Mathematics, University of Manchester, Oxford Road, Manchester, M13
9PL, UK
\\\vspace{6pt}\received{} }

\maketitle

\begin{abstract}
We demonstrate that a quasi-uniform cosmological seed field is a
much less suitable seed for a galactic dynamo than has often been
believed. The age of the Universe is insufficient for a conventional
galactic dynamo to generate a contemporary galactic magnetic field
starting from such a seed, accepting conventional estimates for
physical quantities. We discuss modifications to the scenario for
the evolution of galactic magnetic fields implied by this result. We
also consider briefly the implications of a dynamo number that is
significantly larger than that given by conventional estimates.
\end{abstract}
\bigskip

\bigskip

\section{Introduction}

We start from the concept that galactic magnetic fields are  the result
of dynamo action driven
by galactic differential rotation and mirror asymmetric interstellar
turbulence. This immediately raises the question of the origin and
nature of the seed field from which the
mature field arises. Two options are apparent.

A very weak magnetic field
can be created in a protogalaxy by a battery effect, and then
excited by turbulent motions (the small scale dynamo), to produce a
small-scale field of strength up to approximate equipartition with
the turbulent motions. This then serves as a seed for a mean-field
galactic dynamo, e.g. \cite{b96,r06}. This scenario appears
uncontroversial given our present understanding of
dynamo theory and the physics of galaxies.

Another possibility is that the seed field for the galactic dynamo has
survived from a very early time in the Universe,
see for example the review by \cite{ss05}.
This picture also looks attractive, especially given recent claims
that a significant magnetic field exists in the
intergalactic medium (\cite{ns09}); such a field has been
discussed earlier in various
other contexts, see e.g. \cite{kz08}.
This scenario raises the question of how to form such a magnetic field
of suitable scale, and then to
maintain it against Ohmic losses between the epochs of
recombination and galaxy
formation. The most straightforward possibility perhaps is to consider
a strictly
homogeneous magnetic field which exists from the epoch of the Big Bang, see
\cite{z65}. It has been widely accepted that a galactic mean-field dynamo can,
without any obvious problems, use this field (provided it is sufficiently
strong) as a seed from which to produce contemporary galactic magnetic fields.
This idea has been accepted as obvious in many reviews, e.g.
\cite{b96}.

Observational tests to distinguish between these possibilities are
difficult if not impossible currently. However it may be feasible to
formulate observable predictions from each of them, for
confrontation  with future observations with forthcoming telescopes
such as LOFAR and SKA. Some predictions for tests with  LOFAR were
already suggested in \cite{a09}, \cite{metal2011}.

Our original intention when starting this work was to suggest some observational
tests for the second scenario. In the spirit of this intention we
performed numerical simulations of a galactic dynamo model with a
homogeneous seed. Contrary to the usual expectations, we arrived at
the conclusion that a homogeneous seed may not be as suitable as has been widely
thought for generation of a contemporary galactic magnetic field.
The aim of this paper is to present and develop this finding.

A preliminary description of this work appears in Sokoloff \& Moss
(2011); here we elaborate and expand our investigation.

\begin{figure*}
\begin{center}
\begin{tabular}{ll}
\label{t_evol}
(a)\includegraphics[width=0.44\textwidth]{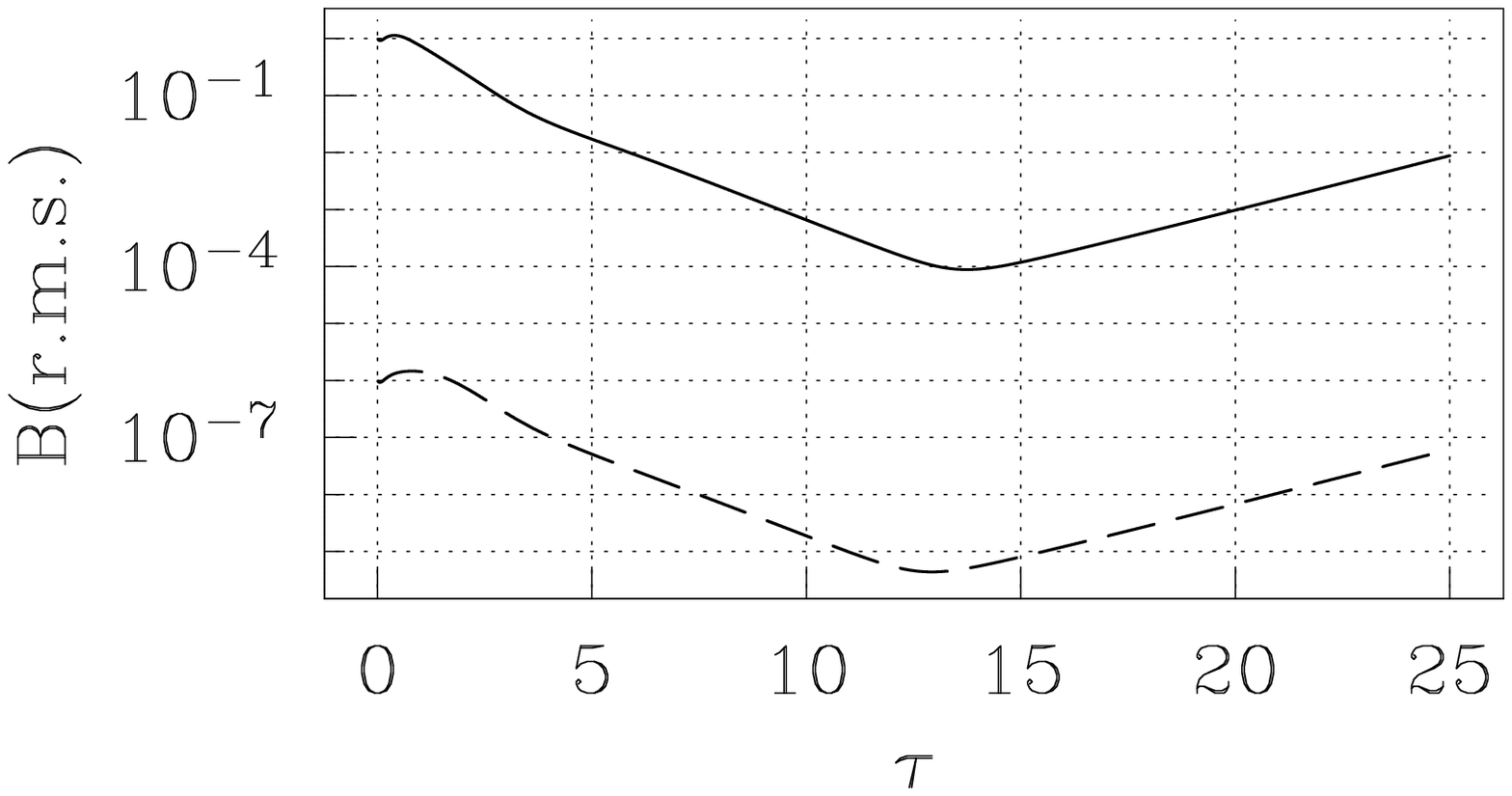} &
(b)\includegraphics[width=0.44\textwidth]{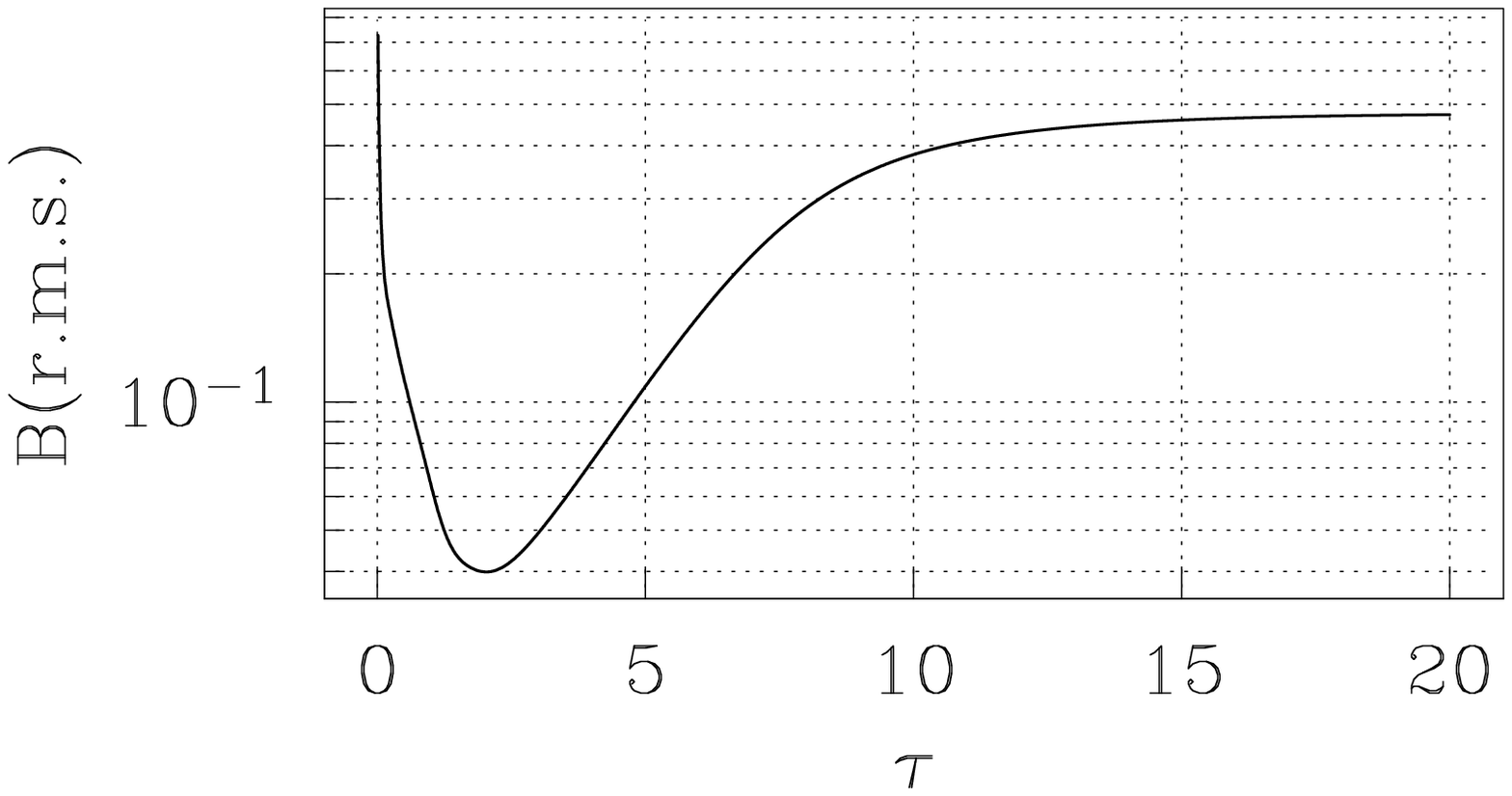}
\\
\end{tabular}
\end{center}
\caption{The evolution of the magnetic field for $R_\alpha=0.8,
R_\omega=8$: (a) uniform seed field with $B_x(t=0)=1$,
alpha-quenched shown as solid, $B_x(t=0)=10^{-6}$, linear by the
broken curve; (b; adapted from \cite{sm11}) random seed field,
$B{\rm r.m.s}(t=0)=1$, alpha-quenched calculation }
\end{figure*}

\section{The code}
\label{code}

Our numerical method has already been described in \cite{sm11},
but for completeness and because our earlier paper may not be easily
accessible, we reproduce here the essential aspects.

We use a thin disc galaxy code that employs the "no-$z$"
formulation, e.g. \cite{sm93,m95} and subsequent papers, taking the
$\alpha\omega$-approximation, and including the correction factors
for the vertical diffusion described in \cite{p01}. This code solves
explicitly for the field components parallel to the disc plane with
the implicit understanding that the component perpendicular to this
plane is given by the condition $\nabla \cdot {\bf B}=0$. The only
novel feature is that the code is now written in cartesian
coordinates $x, y, z$, with $z$-axis parallel to the rotation axis.
(We used a version of the code written in cartesian coordinates,
rather than the perhaps more obvious plane polars, as this had been
found necessary for implementing the ideas described in Moss {\it et
al.} (2011), and is the more sophisticated version.) The key
parameters are the aspect ratio $\lambda=h/R$, where $h$ is the disc
semi-thickness and $R$ its radius, and the dynamo numbers
$R_\alpha=\alpha_0 R/\eta, R_\omega=\Omega_0 R^2/\eta$. $\eta$ is
the turbulent diffusivity, assumed uniform, and $\alpha_0, \Omega_0$
are typical values of  the $\alpha$-coefficient and angular velocity
respectively. Of course, in principle in the $\alpha\omega$
approximation the parameters $R_\alpha, R_\omega$ can be combined
into a single dynamo number $D=R_\alpha R_\omega$, but for reasons
of convenience and clarity of interpretation we choose to keep them
separate. A naive $\alpha$-quenching nonlinearity is assumed in some
of the calculations with i.e. $\alpha=\alpha_0/(1+B({\bf r})^2)$,
where $B$ is the magnitude of the local magnetic field. We
appreciate that there are a variety of other dynamo saturation
mechanisms based on helicity flux mechanisms in the literature, e.g.
\cite{sssb}, \cite{ssub}. It is not the aim of this work to make
detailed comparisons, and indeed most of our conclusions can be
based on strictly linear models because magnetic field remains
safely below the equipartition level, see Fig.~1a. We do however
note that at least for one particular helicity flux mechanism, it
has been shown that alpha-quenching can reproduce the results
reasonably well (Kleeorin et al. 2002).

For the rotation curve we take
\begin{equation}
r\frac{d\Omega}{dr}=-\frac{1}{r R}\tanh(\frac{r R
}{0.3})+\frac{1}{0.3\cosh^2(r R/0.3)}. \label{rot}
\end{equation}
$\alpha=\alpha_0$ is assumed to be uniform.

In order to implement boundary conditions that ${\bf B }\rightarrow
{\bf 0}$ at the disc boundary (as used in earlier versions of the
code written in polar coordinates), the disc region $x^2+y^2\le 1$
was embedded in a larger computational region $|x|\le x_{\rm b},
|y|\le y_{\rm b}$, where $x_{\rm b}=y_{\rm b}\approx 1.35$ ($x$ and
$y$ are measured in units of $R$). In the region $x^2+y^2> 1$
outside of the disc the field satisfies the diffusion equation
without dynamo terms. This was found to give solutions that were
small at the disc boundary, and rapidly became negligible outside of
the disc. The standard implementation used a uniform grid of
$115\times 115$ points over $-1 \le x,y \le +1$, with appropriate
additional points in the surrounding "buffer zone". With a nominal
galaxy radius of 10 kpc, this gives a spatial resolution of about 175 pc.

\section{Computational results}
\label{res}

In this section, we first discuss the temporal evolution of the
magnetic field with uniform and random magnetic fields, as was
already done in \cite{sm11}. Next, we discuss in detail the
differences in the magnetic field configurations in these case. As
in our earlier paper, length, time,  and magnetic field are
nondimensionalized in units of $R$, $h^2/\eta$ and the equipartition
field strength $B_{\rm eq}$ respectively. Taking typical galactic
values, we can estimate $R_\alpha=O(1)$, $R_\omega=O(10)$ (i.e.
$D\approx 10$) with, however, a considerable degree of uncertainty.
Of course these estimates apply to conditions in contemporary spiral
galaxies, and may need revision for very young objects -- we will
return to this later. With $R=10$ kpc, $h=500$ pc gives
$\lambda=0.05$, and the unit of time is then approximately 0.78 Gyr.

From experiments with an arbitrary seed field, the marginal values
are $R_\alpha\approx 0.6$ with $R_\omega=8$, so the above estimates
correspond to a slightly supercritical dynamo.

We first took a homogeneous seed field, $B_x=1, B_y=0$, with the
slightly supercritical values $R_\alpha=0.8, R_\omega=8$. The field
is wound up by the differential rotation in the anticipated manner,
but its strength decays until time $\tau>14$, i.e. of order the age
of the universe -- see the evolution of the field strength with time
given by the solid curve in Fig.~1a. In this computation the alpha
quenching nonlinearity was implemented, but it rapidly becomes
irrelevant as the field strength decreases. This is illustrated by
the broken curve in Fig.~1a which summarizes a linear computation
with initial $B_x=10^{-6}, B_y=0$. The two curves are essentially
identical but displaced by $10^{-6}$ in $B{\rm (r.m.s.)}$ values.
Only when the dynamo numbers are increased to something like
$R_\alpha=0.8, R_\omega=20$ is there unambiguous growth after a
relatively short initial relaxation phase. Note also that the
observational upper bound for the homogeneous cosmological magnetic
field is safely below equipartition, see e.g. \cite{b96}.

\begin{figure*}
\begin{center}
\vspace{0.5cm}
\begin{tabular}{ll}
\label{B_evol}
\includegraphics[width=0.25\textwidth]{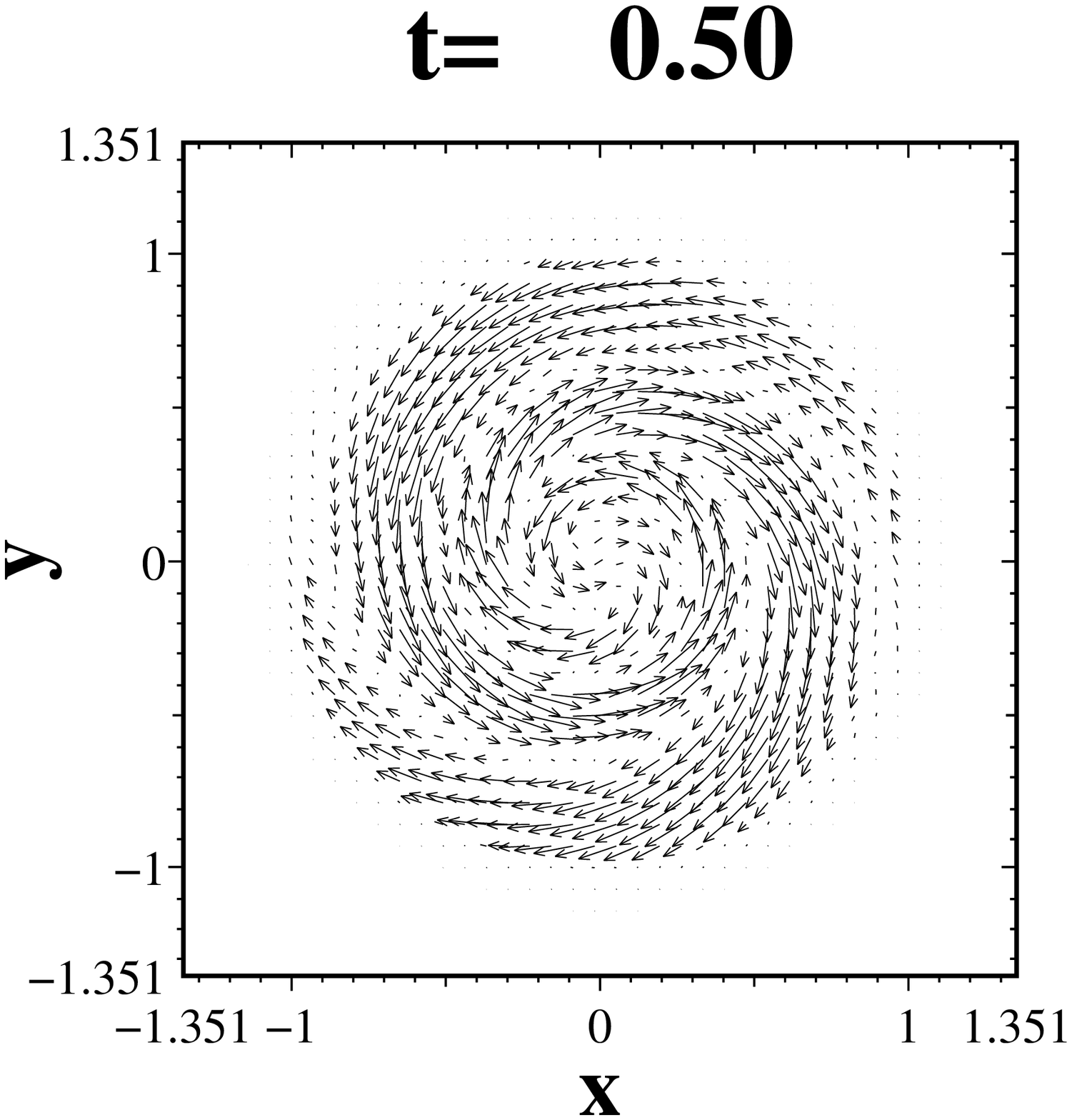} &
\includegraphics[width=0.25\textwidth]{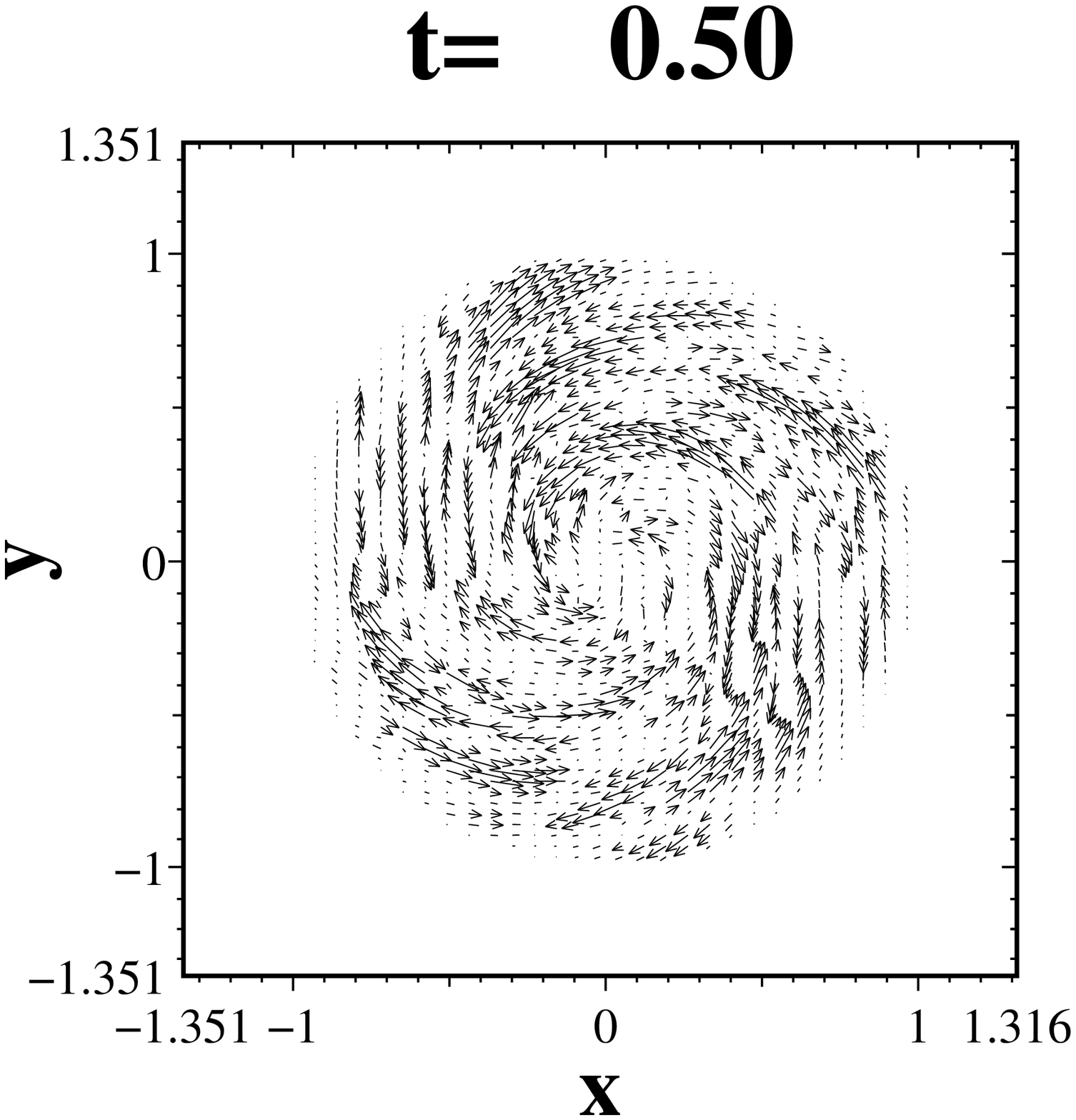} \cr
\includegraphics[width=0.25\textwidth]{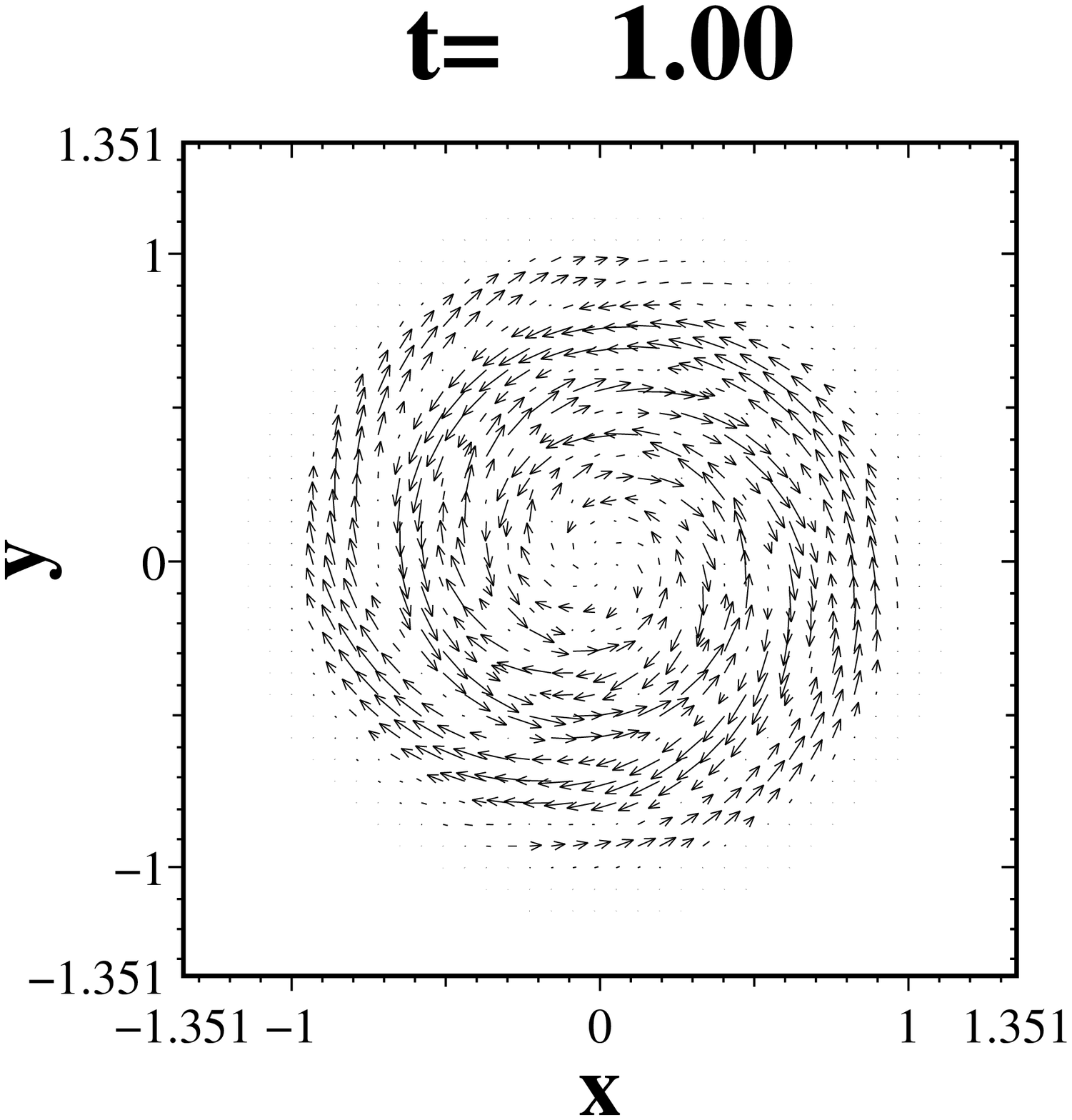} &
\includegraphics[width=0.25\textwidth]{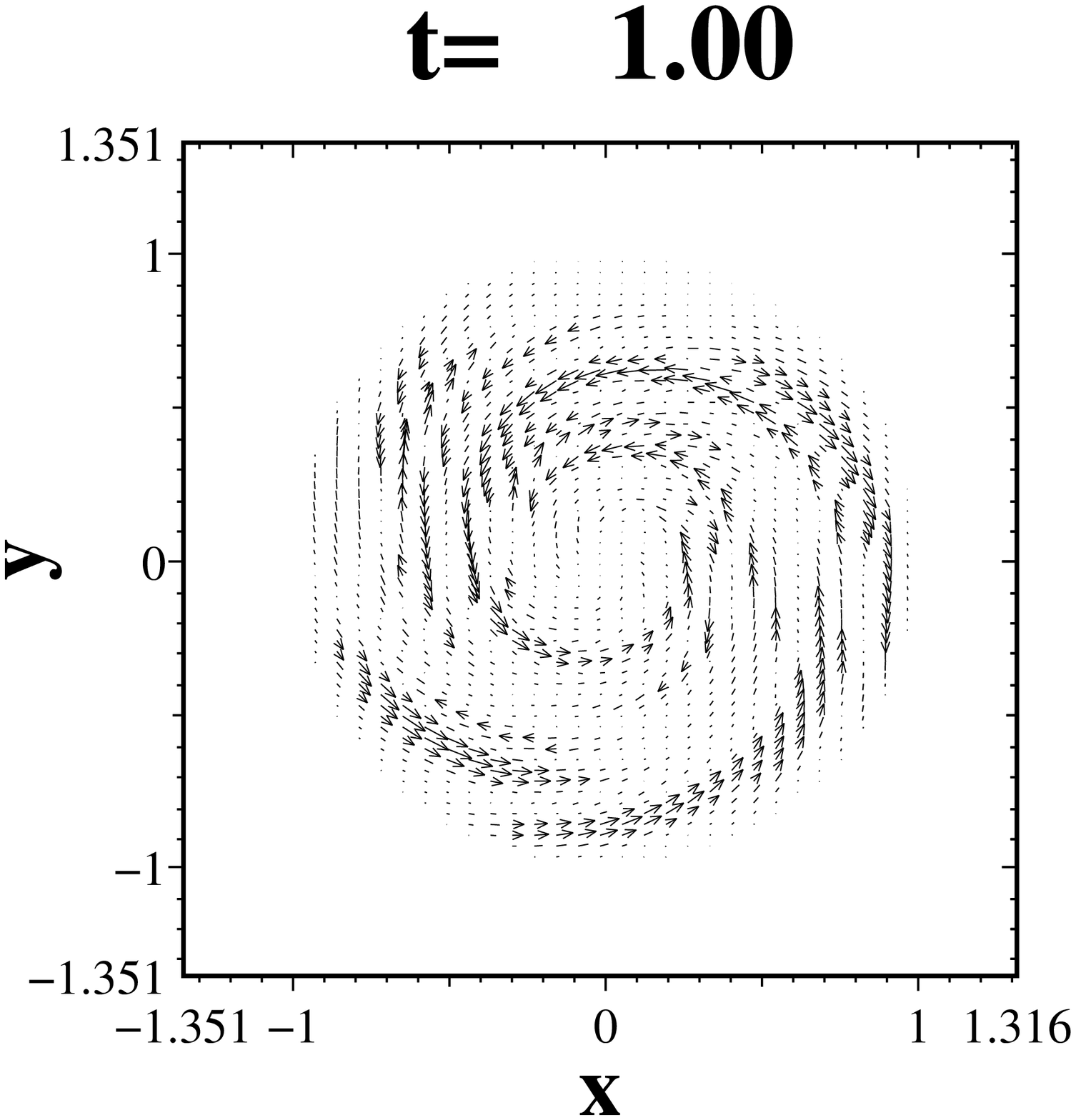} \cr
\includegraphics[width=0.25\textwidth]{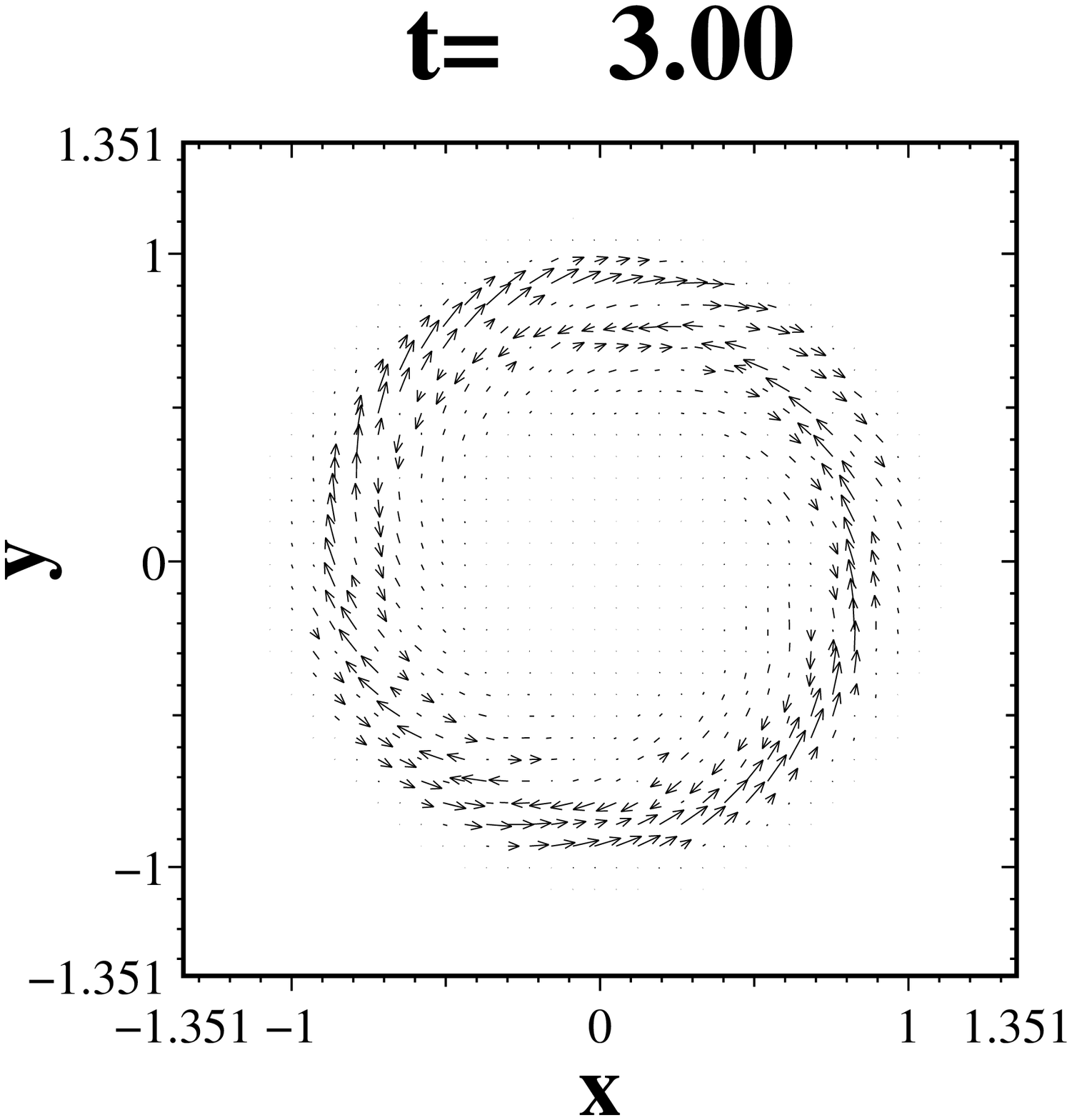} &
\includegraphics[width=0.25\textwidth]{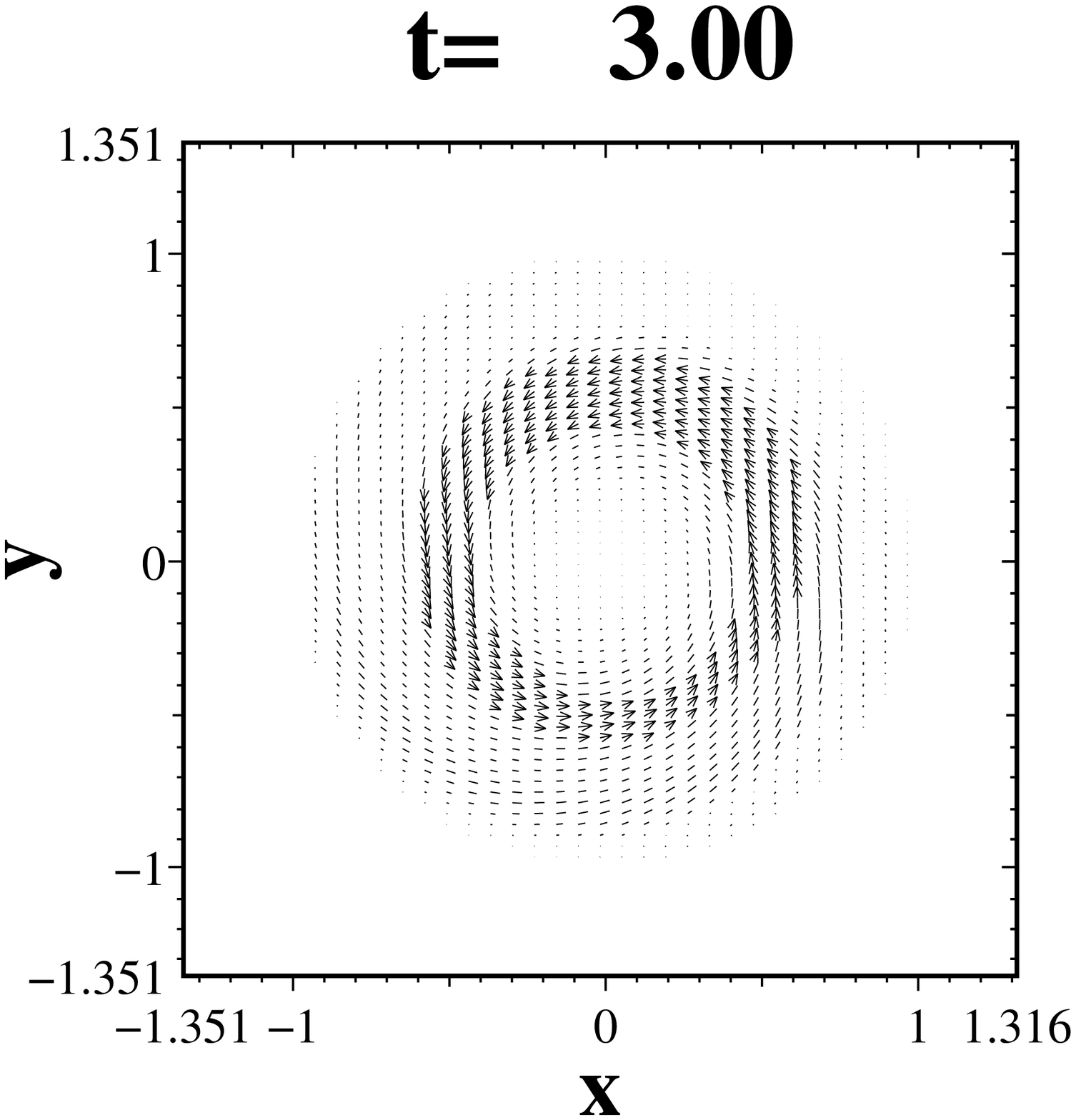} \cr
\includegraphics[width=0.25\textwidth]{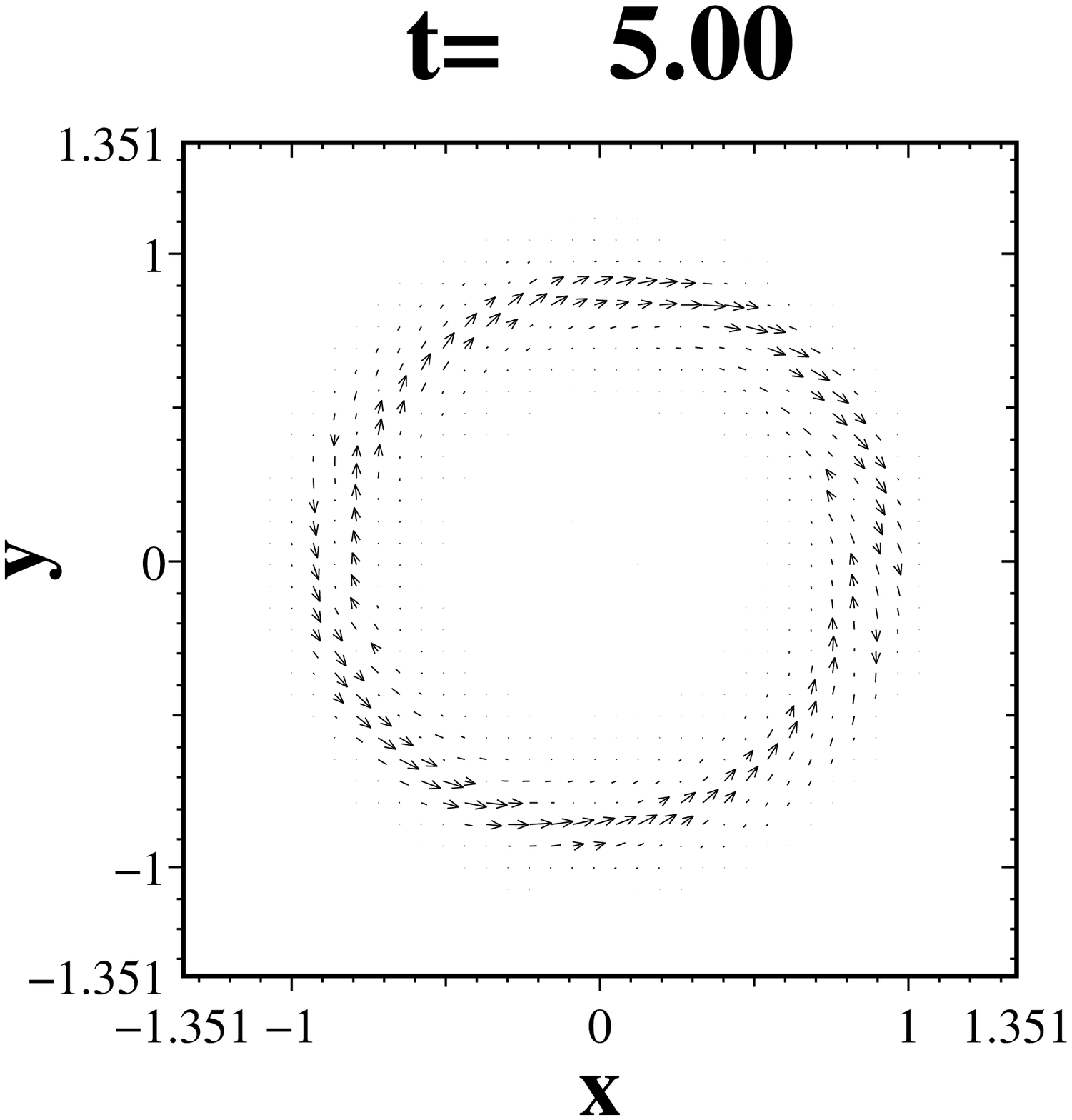}&
\includegraphics[width=0.25\textwidth]{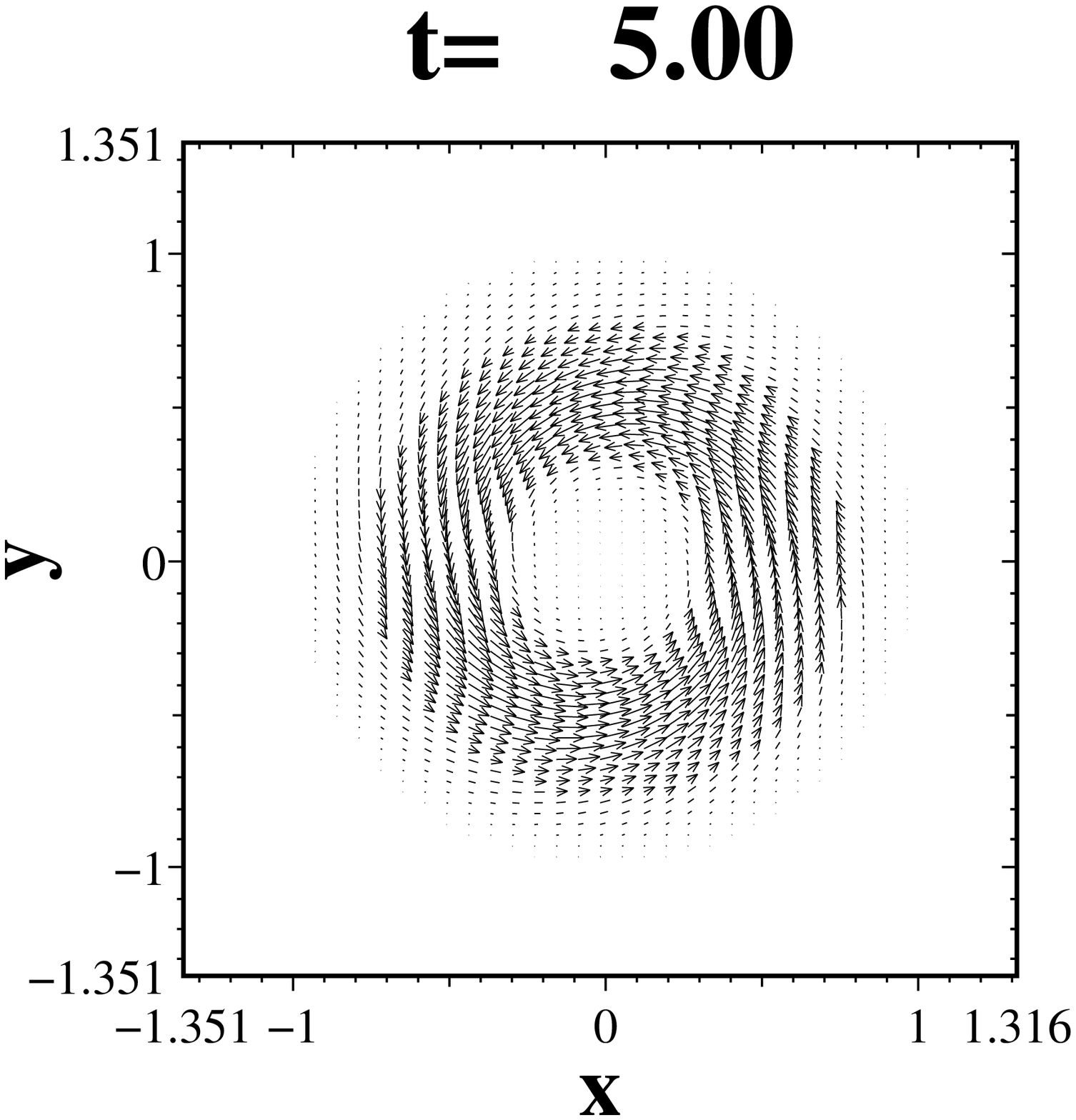}\cr
\includegraphics[width=0.25\textwidth]{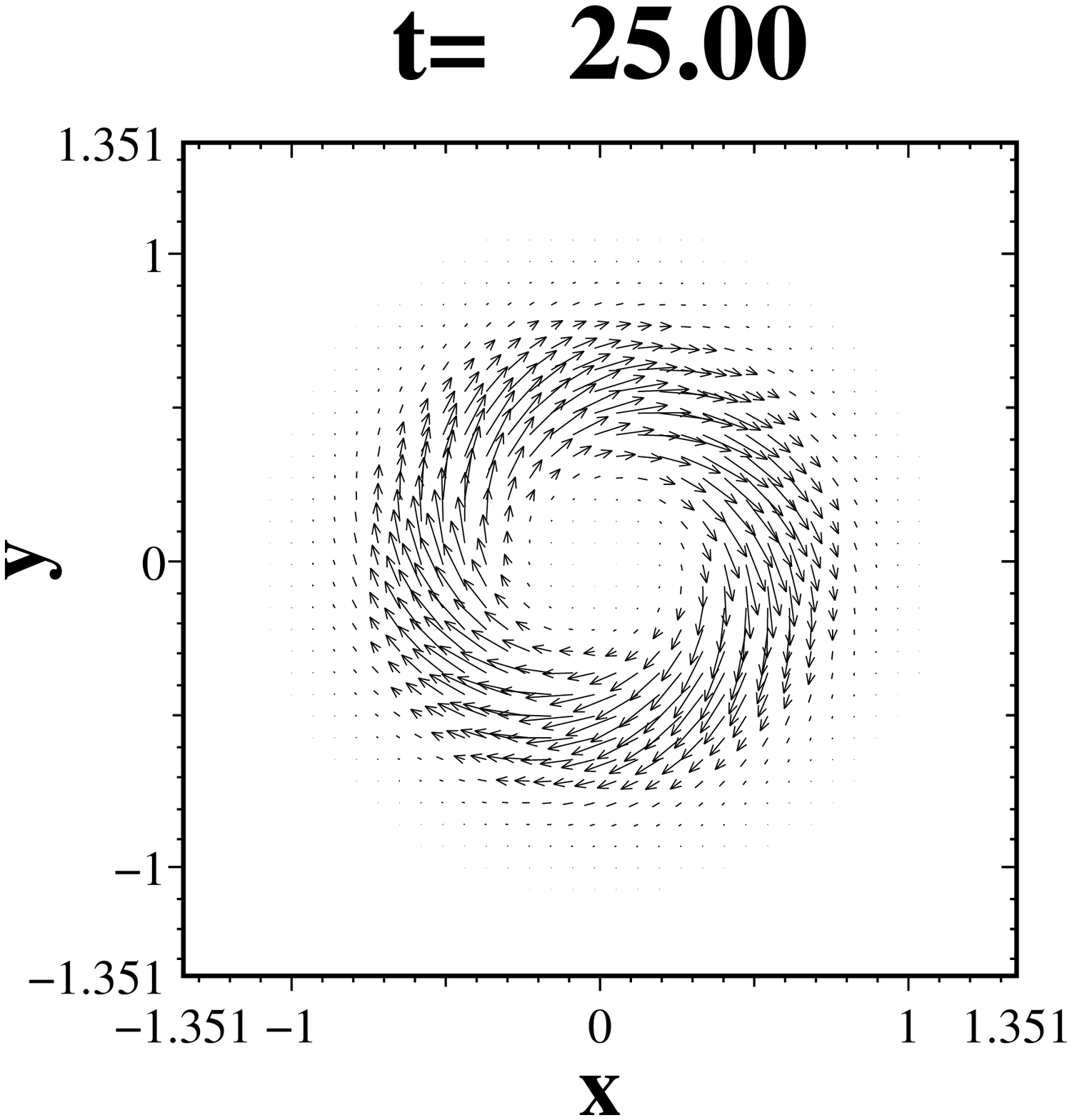} &
\\
\end{tabular}
\end{center}
\caption{Evolution of horizontal magnetic field for a
homogeneous seed field (left
hand column) and random seed (right hand column). The 2D components of magnetic
fields in the equatorial plane are shown. Dimensionless times are indicated
above the plots. We stress that scenario shown in the left hand column takes
much more time to reach a more or less steady configuration than that
shown on the right. Indeed, the magnetic energy of the field from the
homogeneous seed is, at $t=25$, still growing -- see Fig.~1a.}
\end{figure*}

\begin{figure*}
\begin{center}
\vspace{0.5cm}
\begin{tabular}{llll}
\label{t_evol2}
\includegraphics[width=0.23\textwidth]{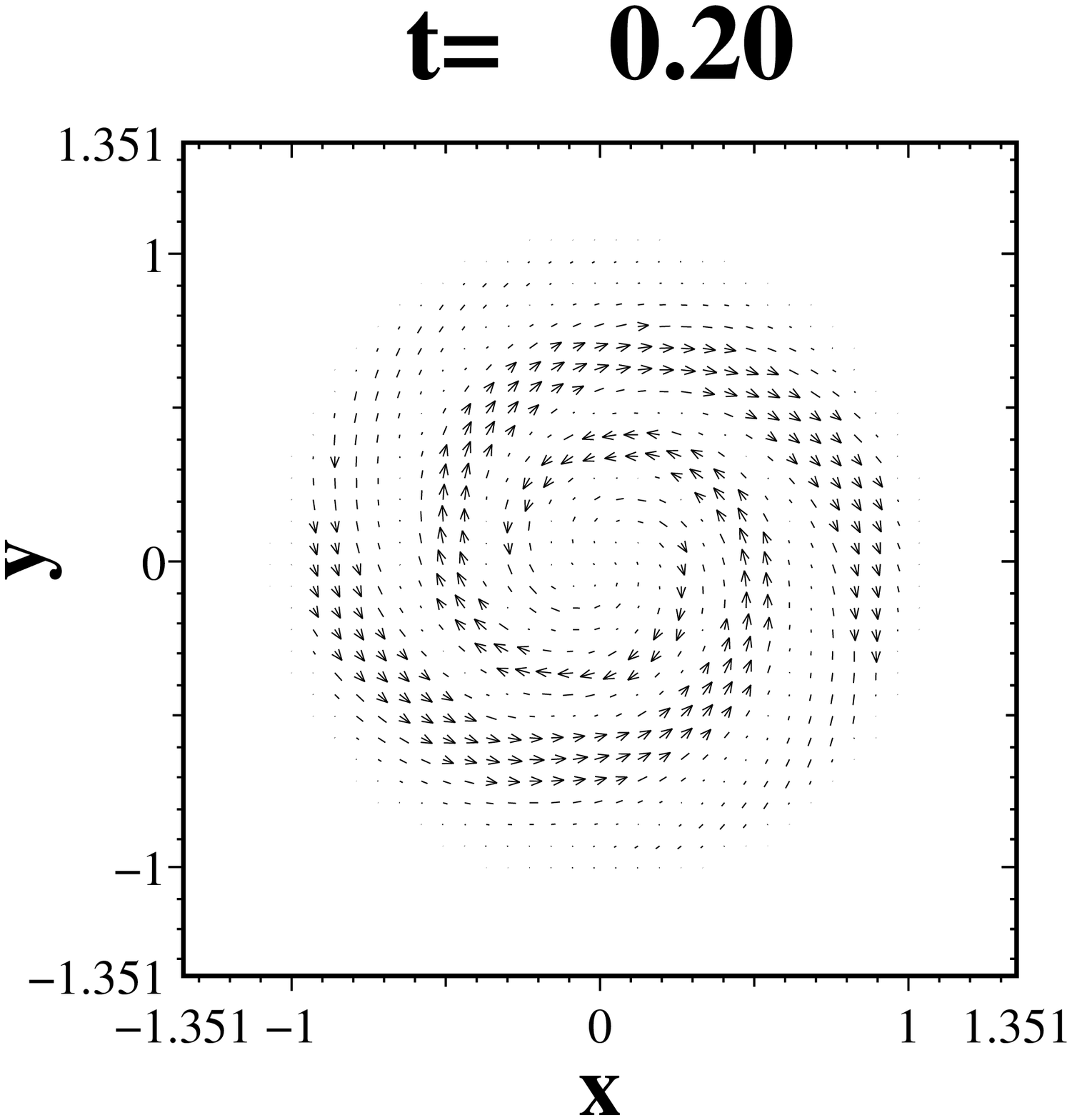} &
\includegraphics[width=0.23\textwidth]{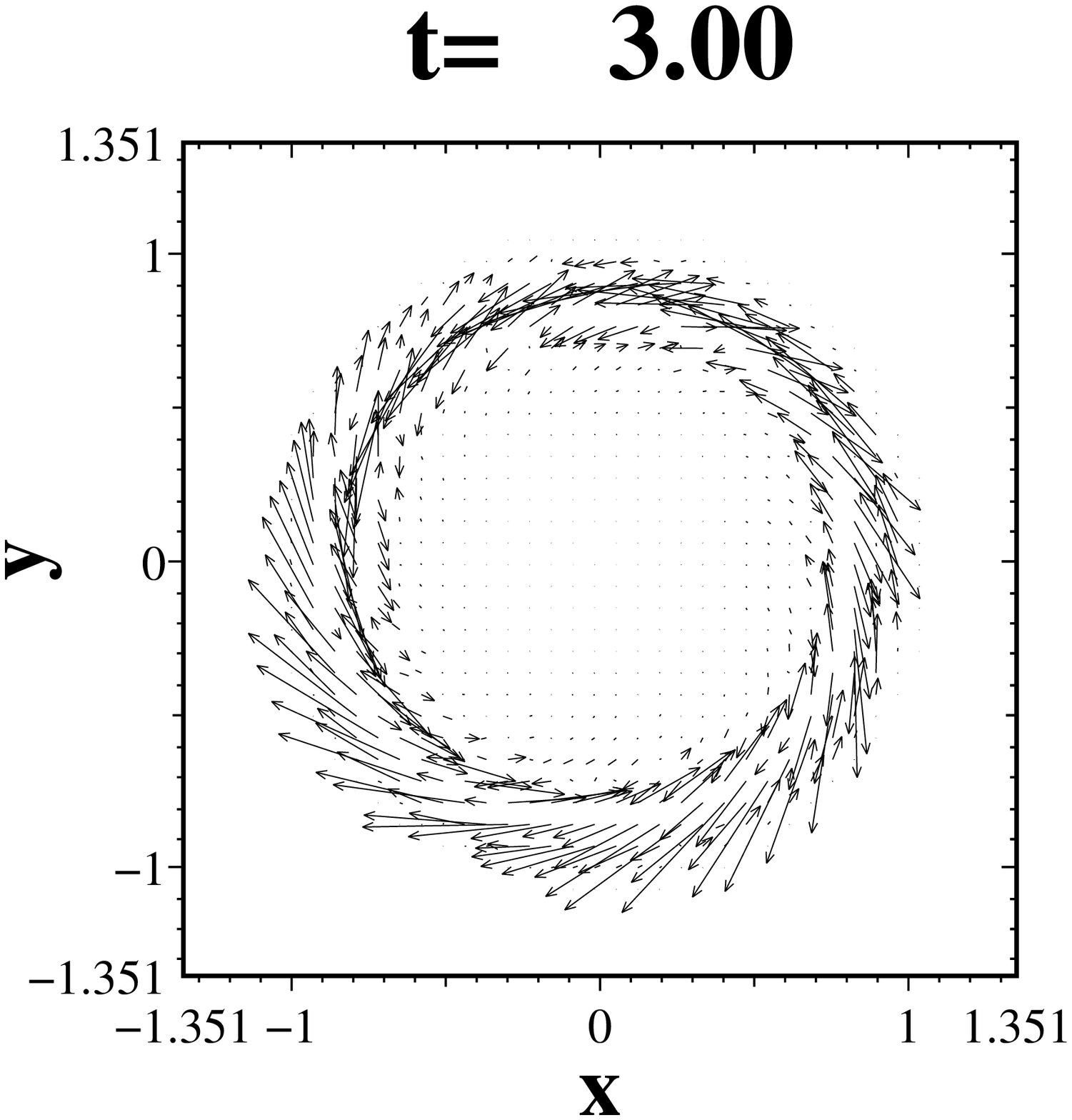} &
\includegraphics[width=0.23\textwidth]{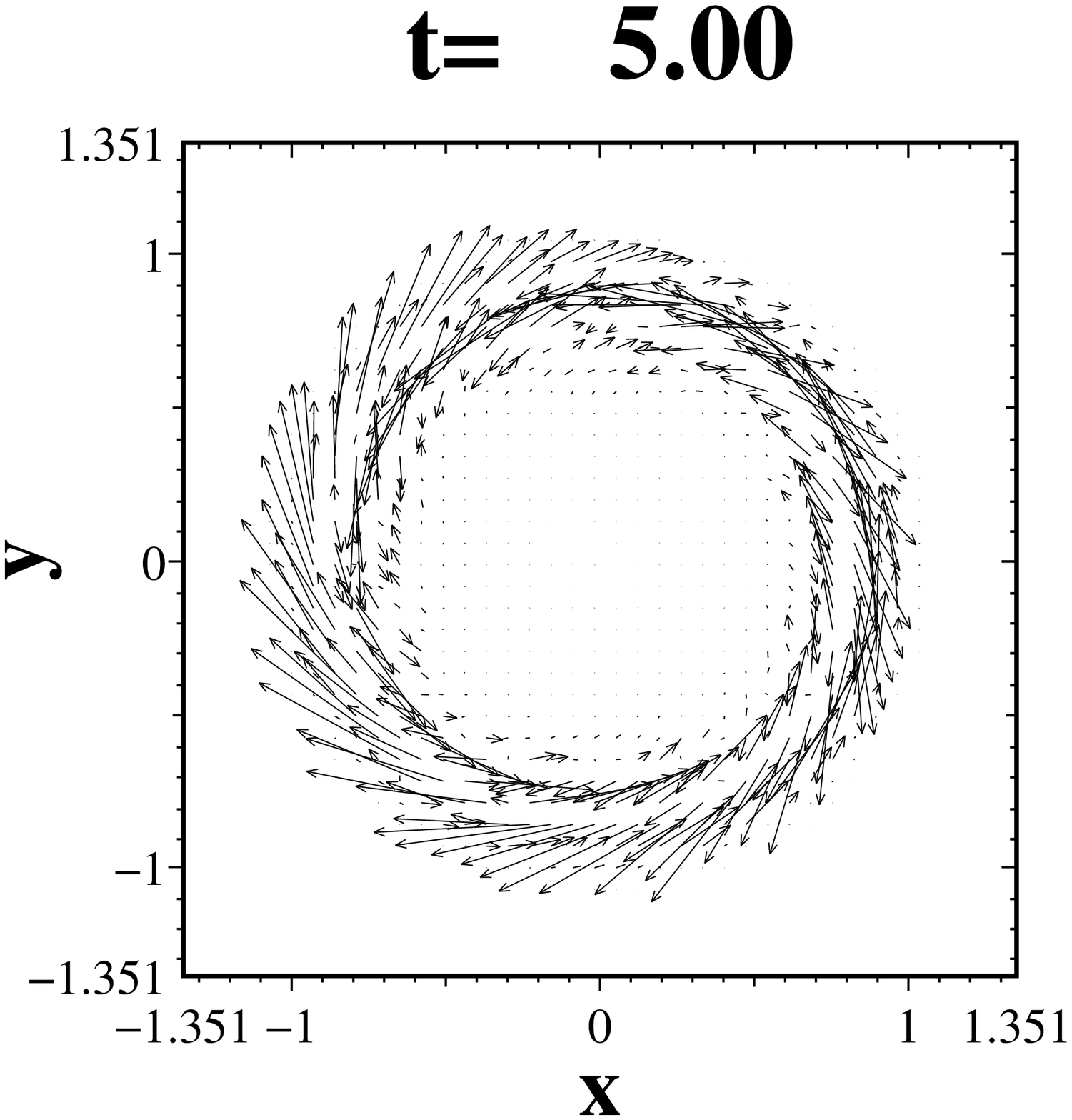} &
\includegraphics[width=0.23\textwidth]{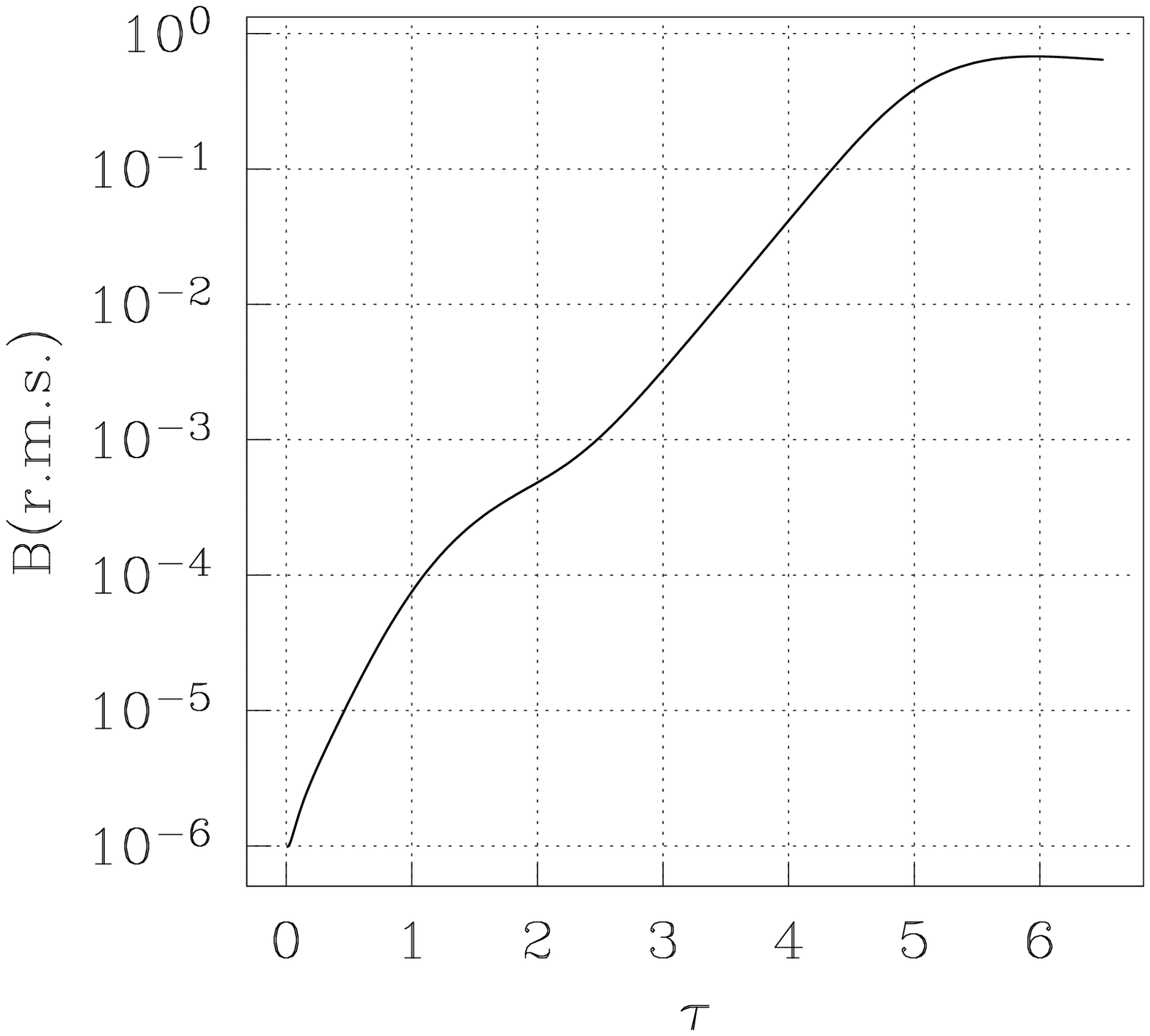}
\\
\end{tabular}
\end{center}
\caption{Magnetic field evolution for strong dynamo action.
First three panels: evolution of horizontal magnetic field for a
strong homogeneous seed field with dynamo parameters $R_\alpha=1,
R_\omega=20$, with alpha-quenching. Dimensionless times are shown
above the plots. Panel 4: saturation with typical field strengths
about unity is reached by time $t=5$.}
\end{figure*}

In contrast to this very ordered seed field, we then took a field
that was random in $B_x, B_y$ on the scale of the grid, with r.m.s.
value of unity. Now, with $R_\alpha=0.8, R_\omega=8$ growth is seen
after a short initial relaxation - see Fig~1b.

The explanation for these contrasting behaviours seems clear and
straightforward. The uniform seed field has a Fourier component
$m=1$ only. In contrast, the random seed field contains all Fourier
modes $m$. It is well known that it is harder to excite an $m=1$
field than $m=0$, e.g. \cite{rss88}. In the second example the $m=0$
part of the seed is selected and amplified whilst, while in the
linear regime at least, the other components decay. In Fig.~1a the
eventual growth of the homogeneous seed when $\tau\gta 14$ is from
the amplification of numerical noise. (When this case is rerun with
a code written in polar coordinates, this growth appears at a
somewhat later time: the polar code preserves the Fourier structure
of the field rather better than the cartesian.)

In Fig.~2 we show the temporal evolution of the magnetic field for a
random seed magnetic field (right hand column) and a homogeneous
seed magnetic field (left hand column) -- they are quite different.
We see from the left hand column that the initial homogeneous seed
field is first almost destroyed by differential rotation and later
(after $t \approx 5.0$) a ring of axisymmetric magnetic field
arises. In contrast, magnetic spots in the right hand column
agglomerate soon after time $t \approx 1$ to yield in an
axisymmetric magnetic field. In the  first case, growth takes much
more time than in the second.

In several other simulations with $R_\alpha=0.8, R_\omega=8$ and
different large-scale seed fields we found that those with no $m=0$
component behaved much as shown in Fig.~1a, whereas the behaviour of
those with a $m=0$ part resembled more closely that of Fig.~1b, with
the initial fall in $B{\rm (r.m.s.)}$ being smaller for larger
dynamo numbers or a larger proportion of the energy in the $m=0$
component. To illustrate this point more clearly, we show in
the first three panels of Fig.~3
snapshots of field evolution with $R_\alpha=1, R_\omega=20$ and an
initially homogeneous field of magnitude $10^{-6}$. Alpha-quenching
is implemented for better comparison with previous results; the
field grows steadily and saturates with mean strength around unity
by time $t=5$ -- see the 4th panel in Fig.~3.

\section{Discussion and conclusions}

We see that we have come to a rather unexpected conclusion: a
cosmological magnetic field that is uniform on the scale of the
young galaxy is a "bad" seed for mean-field
galactic dynamos.

Of course, this conclusion holds for the conventional estimate for
the intensity of dynamo action, as measured by the dynamo number
$D$, and estimated using knowledge of contemporary galaxies. It is
possible that this estimate should be reconsidered for very young
galaxies -- see e.g. the example illustrated in Fig.~3. In
principle, this possibility might be pursued to attempt to produce bisymmetric
magnetic configurations in galaxies. However, the situation is
unclear, and galactic dynamos can anyway work without any
cosmological seed field.

One possibility is that a relatively strong initial field (whose
origin we are not considering) undergoes dynamical instabilities,
which generate small-scale magnetic fields which contain
nonvanishing $m=0$ components, and these then grow in a similar
manner to that shown in Fig.~1b and the right hand column of Fig.~2.

Related in some ways to this is the possibility that a uniform field,
being affected by, say,
small-scale turbulence,
acquires all modes including $m=0$. In either of these cases a
mean-field dynamo can take as a seed the $m=0$ component so generated,
and amplify
it up to the currently observed field strengths and configurations.
Thus our
result does not exclude in principle the role of cosmological seed  fields.
The point however is that the growth of the initial field component
$m=0$ is then determined by a particular distortion that creates it,
rather than being present in the cosmological field itself. Numerous
such distortions will  occur during early galactic evolution, and so
the real difference between two scenarios seems to be much less
important than is usually assumed.

Our result is valid for both strong and weak initial fields, and the
inclusion of a simple saturation mechanism makes essentially no
difference; with a strong seed  after a short initial period the
evolution is essentially linear.

Our interpretation of the results obtained needs however some comments. The no-$z$ model
for a strictly homogeneous
seed (i.e. uniform, in any direction) describes the magnetic field component perpendicular to the rotation axis
while the component parallel
to the rotation axis remains unaffected by the rotation.

If there is a large-scale homogeneous seed of cosmological (or
very early pre-galaxy formation) origin then the component parallel
to $\Omega$ is of aligned dipolar parity, and the perpendicular component
has the parity of a perpendicular dipole (i.e. $m=1$).
The no-$z$ model can handle the latter case (but not the former,
being explicitly restricted to fields of even parity with respect
to the galaxy mid-plane).
It shows that it is a "bad" seed field in sense that corresponding dynamo
generated magnetic field develops too slowly to
generate the present-day galactic magnetic field.
The  code can also handle a axisymmetric quadrupole-like seed,
which grows normally (seed is, e.g., purely radial in the no-$z$ formulation).
Classical axisymmetric disc dynamo models
show that axisymmetric  fields of quadrupolar parity are preferred over
those of dipolar parity,
 e.g. \cite{rss88} and much subsequent work.
Thus we deduce that large scale homogeneous seeds
are inefficient, whatever their orientation -- large-scale quasi-homogeneous seed fields can hardly have quadrupolar parity.

In this context, another limitation of the no-$z$ model to be
considered is that it cannot explicitly include the effects of
galactic winds blowing out of the disc, or of turbulent diamagnetism
(in some ways these have opposing effects). Both of these mechanisms
were investigated by Brandenburg {\it et al.} (1993), who found that
solutions with quadrupolar parity remained preferred -- consistent
with a basic assumption of the no-$z$ model. \cite{metal10}
similarly demonstrated that the preference for quadrupolar fields
persisted under a wide range of conditions; globally mixed parity
solutions are also possible, although these are usually dominantly
quadrupolar in the disc region.

We recognize that widely discussed mechanisms for magnetic field
generation in the early universe give rise to small-scale fields. In
such cases, the subsequent evolution of galactic fields can not
expect to be distinguished from that following small-scale dynamo
action in the proto-galaxy. In this paper we have mainly considered
the consequences of an alternative scenario (which we do not
particularly advocate), that a large-scale quasi-homogeneous
cosmological field is present at the time of galaxy formation.

Note also that contemporary dynamo theory considers dynamo models
that are much more detailed than is used by our simple mean-field
dynamo model, e.g. \cite{hetal09}, \cite{kd11}, 
\cite{getal08}. Models based on direct numerical simulations of the
induction equation rather than on mean-field dynamo equations,
include however both the galactic dynamo action and distortion of
the uniform field by small-scale motions simultaneously, and
demonstrate the joint action of both effects. In contrast,  in this
paper we have separated them, in order to elucidate the basic
principles.

We are grateful to our colleagues in research on magnetic field
evolution in very young galaxies, T. Arshakian, R. Beck, M. Krause,
R. Stepanov, for fruitful discussions which stimulated our
writing this the paper. DS acknowledges financial support from NORDITA for his participation in the meeting.

\end{document}